
%
\let\includefigures=\iftrue
%
%
\let\useblackboard=\iftrue
%
%
%
\let\expandedversion=\iffalse
\input harvmac.tex
\includefigures
\message{If you do not have epsf.tex (to include figures),}
\message{change the option at the top of the tex file.}
\input epsf
\epsfclipon
\def\fig#1#2{\topinsert\epsffile{#1}\noindent{#2}\endinsert}
\else
\def\fig#1#2{\vskip .5in
\centerline{Figure 1}
\vskip .5in}
\fi
\def\Title#1#2{\rightline{#1}
\ifx\answ\bigans\nopagenumbers\pageno0\vskip1in%
\baselineskip 15pt plus 1pt minus 1pt
\else
\def\listrefs{\footatend\vskip 1in\immediate\closeout\rfile\writestoppt
\baselineskip=14pt\centerline{{\bf References}}\bigskip{\frenchspacing%
\parindent=20pt\escapechar=` \input
refs.tmp\vfill\eject}\nonfrenchspacing}
\pageno1\vskip.8in\fi \centerline{\titlefont #2}\vskip .5in}

\ifx\answ\bigans\def\tcbreak#1{}\else\def\tcbreak#1{\cr&{#1}}\fi
\useblackboard
\message{If you do not have msbm (blackboard bold) fonts,}
\message{change the option at the top of the tex file.}
\font\blackboard=msbm10 scaled \magstep1
\font\blackboards=msbm7
\font\blackboardss=msbm5
\newfam\black
\textfont\black=\blackboard
\scriptfont\black=\blackboards
\scriptscriptfont\black=\blackboardss

\else

\fi
%
\def\yboxit#1#2{\vbox{\hrule height #1 \hbox{\vrule width #1
\vbox{#2}\vrule width #1 }\hrule height #1 }}
\def\fillbox#1{\hbox to #1{\vbox to #1{\vfil}\hfil}}
\def\ybox{{\lower 1.3pt \yboxit{0.4pt}{\fillbox{8pt}}\hskip-0.2pt}}
\def\comments#1{}

\def\p{\partial}
\def\eps{\epsilon}

\def\Tr{{\rm Tr\ }}
\def\tr{{\rm tr\ }}

\def\im{{\rm Im\hskip0.1em}}
\def\bra#1{{\langle}#1|}
\def\ket#1{|#1\rangle}
\def\vev#1{\langle{#1}\rangle}

\def\CF{{\cal F}}

\def\CS{{\cal S}}

\def\Cr{\hfill\break}
\def\a{\alpha}
\Title{\vbox{\baselineskip12pt
\hfill{\vbox{
\hbox{RU-94-81\hfil}
}}}}
{\vbox{\centerline{Stochastic Master Fields}}}
\centerline{Michael R. Douglas}
\smallskip
\centerline{Dept. of Physics and Astronomy}
\centerline{Rutgers University }
\centerline{\tt mrd@physics.rutgers.edu}
\bigskip
\bigskip
\noindent
We treat the stochastic equation for large $N$ master fields proposed by
Greensite and Halpern using a
construction of master fields modelled after work
of Voiculescu, and show that it contains the same information as the usual
factorized Schwinger-Dyson equations.\Cr
We comment on the relation to earlier work of Haan and of
Cvitanovic, Lauwers and Scharbach.
\Date{November 1994}
\nref\voi{D.~V.~Voiculescu, K. J. Dykema and A. Nica, {\it Free Random
Variables}, AMS 1992.}
\nref\singer{I.~Singer, ``The Master Field for Two-Dimensional Yang-Mills
theory,'' lecture at the July 1994 Mathematical Physics Conference in Paris,
and to appear.}
\nref\doug{M.~R.~Douglas, ``Large N Gauge Theory -- Expansions and
Transitions,'' May 1994 lectures at the ICTP, hep-th/9409098.}
\nref\greenhal{J.~Greensite and M.~B.~Halpern, Nucl.Phys. B211 (1983) 343.}
\nref\bardakci{K.~Bardakci, Nucl.Phys. B219 (1983) 302.}
\nref\jevicki{A.~Jevicki and H.~Levine, Ann.Phys. 136 (1981) 113.}
\nref\halpern{M.~B.~Halpern, Nucl.Phys. B188 (1981) 61.}
\nref\halpschw{M.~Halpern and C.~Schwartz, Phys.Rev. D24 (1981) 2146.}
\nref\bipz{E. Brezin, C. Itzykson, G. Parisi and J.-B. Zuber,
Comm.Math.Phys 59 (1978) 35.}
\nref\gross{R.~Gopakumar and D.~Gross, private communication.}
\nref\nsf{I.~M.~Gelfand, D.~Krob, A.~Lascoux, B.~Leclerc, V.~S.~Retakh and
J.-Y.~Thibon, preprint LITP94.39, hep-th/9407124.}
\nref\stoquant{{\it Stochastic Quantization,} eds. P.~Damgaard and H. H\"uffel,
World 1988.}
\nref\parisi{G.~Parisi and Wu Yongshi, Sci.Sin. 24 (1981) 483.}
\nref\breit{J.~D.~Breit, S.~Gupta and A.~Zaks, Nucl.Phys. B223 (1984) 61.}
\nref\bhst{Z.~Bern, M.~B.~Halpern, L.~Sadun and C.~Taubes, Phys.Lett. B165
(1985) 151.}
\nref\dosch{See for example H.~G.~Dosch, Phys.Lett. 190B (1987) 177.}
\nref\grosspa{R.~Gopakumar and D.~Gross, Princeton preprint PUPT-1520,\Cr
hep-th/9411021.}
\nref\haan{O.~Haan, Z.Physik C6 (1980) 345.}
\nref\cls{P.~Cvitanovic, P.~G.~Lauwers and P.~N.~Scharbach, Nucl.Phys. B203
(1982) 385.}
\newsec{Introduction}

Recent mathematical work (\refs{\voi,\singer}; an introduction for physicists
is in \doug) has produced constructions
of the `master field' for general large $N$ gauge and matrix field theories.
The essential difficulty of the large number of degrees of freedom in higher
dimensional large $N$ theories is dealt with by finding master fields which
live in `large' operator algebras such as the type II$_1$ factor associated
with a free group.
Singer has advocated these ideas as the appropriate framework for treating the
large $N$ limit of general field theories such as four-dimensional QCD.
\singer

So far the explicit constructions have been for very simple solvable theories
such as decoupled multi-matrix integrals or two-dimensional Yang-Mills theory.
The constructions required as input correlation functions computed using other
techniques (such as summing diagrams or the existing saddle-point methods) and
thus are not explicit for physically more interesting theories.
One way to treat such theories would be to
find equations satisfied by master fields and solve them.
This approach was worked on extensively in the early 80's
(some references are \refs{\greenhal-\halpschw})
and several such equations were proposed, identical to or derived directly from
the
original classical equations of motion.
However, the lack of analytic techniques for
working with master fields stopped progress in this direction.

In this paper we will apply the new ideas in the large $N$ stochastic approach
of Greensite and Halpern.  \greenhal\ %
They proposed interpreting the Langevin equation of stochastic quantization,
\eqn\stoone{{\p\over \p\tau} \phi(\tau,x) =
-{\delta\over\delta\phi(\tau,x)} S[\phi] + \eta(\tau,x)}
as an equation for a master field $\phi$.  The main advantage of large $N$ in
this approach is that one eliminates the second step of stochastic
quantization, averaging over the noise $\eta$, by using a `master noise
source,' a master field $\eta$ reproducing Gaussian correlations.
The claim is then that the proper time evolution has a large $\tau$ fixed point
which is the master field, and Greensite and Halpern showed this in
perturbation theory.

Our main result is to show how this can be made precise using a precise
definition of master field, and to show its equivalence to the standard
factorized Schwinger-Dyson equations.  A complete treatment is given only for a
solvable case (the one-matrix model) but the ideas are general.  Techniques for
solving the equations we discuss in multi-matrix or higher dimensional models
do not exist, but at least one can describe what one is looking for.

Let us make two remarks on similar results which should also exist.  First, the
canonical formalism was also considered \refs{\bardakci,\jevicki,\halpschw}\
and it appears to us that an analogous but even simpler statement
might be made -- that the field $\phi(\vec x)$ and its canonical conjugate
$\Pi(\vec x)$ can be reproduced by master fields which (in some sense) satisfy
the classical equation of motion $\dot\phi=\{H,\phi\}$ with the {\it original}
Poisson bracket.  The simplicity of this is very attractive but a major
unsolved problem in this approach is to characterize {\it which} points in
phase space $(\phi,\Pi)$ are possible master fields (for example, $\phi=\Pi=0$
clearly is not).  Second, many workers and in particular Greensite and Halpern
found it useful to eliminate the space-time dependence of the fields in favor
of structure in the internal variables by the procedures of `reduction' and
`quenching.'  We have little to say about this; it is not even clear to us
whether it is a step forward or backward.

In section 2 we discuss master fields, in section 3 we review stochastic
quantization, and in section 4 give our main results.
In this version, we have commented on the relation to earlier work of Haan
\haan\ and of
Cvitanovic, Lauwers and Scharbach \cls\ in section 5, and in section 6 state
conclusions.

\newsec{Master fields}

We will consider a matrix model with fields $M_\alpha$ with $\alpha\in S$ a
set,
and action $S[M]$.
Most of what we say applies for a field theory if we take $S$ to include
space-time, $\delta_{\alpha,\beta}$ to include a Dirac delta function, and so
forth.

It is worth making a few general comments about master fields.
We can regard the set of all vacuum expectation values
\eqn\vevdef{\vev{\Tr (M_{\alpha_1} M_{\alpha_2}\ldots)} =
\lim_{N\rightarrow\infty}
{1\over N}{1\over Z}\int DM~e^{-NS[M]}~ {\Tr M_{\alpha_1} M_{\alpha_2}\ldots}}
as a linear functional $W(a)$ on the algebra of words generated by formal
variables $m_\alpha$ with $\alpha\in S$.
The functional depends on $S$ and if we consider more than one action, we would
indicate this dependence as $W_S(a)$.
In the large $N$ limit, no kinematic (independent of $S$) equality relations
between expectations of different words are known (the finite $N$ Mandelstam
relations certainly have no obvious limit) and we claim that none exist.  (See
\doug\ for arguments in this direction, which we will pursue elsewhere.)
Thus to represent the general functional $W$, we need to consider variables
$m_\alpha$ satisfying no relations: the algebra is the free algebra on $n=|S|$
(cardinality of $S$) generators; call it $\CF_n$.

The problem of finding a master field is the problem of finding operators $\hat
M_\alpha$ and trace $\hat\Tr$ such that
\eqn\Masdef{\hat\Tr \hat M_{\alpha_1} \hat M_{\alpha_2}\ldots\hat M_{\alpha_k}
= W(m_{\alpha_1} m_{\alpha_2}\ldots m_{\alpha_k}).}
It turns out that the problem cannot be solved using the usual $\Tr M = \sum_i
M_{ii}$ and one must instead consider more general traces satisfying linearity
and $\hat\Tr AB=\hat\Tr BA$ (which implies $\hat\Tr U^+ AU=\hat\Tr A$ for a
unitary $U$).

The solution will only be unique up to unitary equivalence.
For a one matrix integral, the following solution is a perfectly good one:
$\hat M = \lambda$ and $\hat\Tr \hat O = \int d\lambda \rho(\lambda) \hat O$.
We might consider others in which $\hat M$ was not diagonal.
For example, the method of orthogonal polynomials provides a representation
$\hat M = \sum R_n (e_{n,n+1}+e_{n+1,n})$ and
$\hat\Tr \hat O = \sum_{i=1}^N O_{ii}$.
Since the inner product does not enter \Masdef, we might even consider general
similarity transformations $\CS M_i \CS^{-1}$ with $\CS$ any invertible
operator.

The first criterion we must satisfy is that the $\hat M_\alpha$ form a
representation of the algebra $\CF_n$.  At the formal level we work at in this
paper, $\CF_n$ has no structure, and this gives no constraint on the $\hat
M_\alpha$.  As we will see shortly however the $\hat M_\alpha$ are operators on
a very large Hilbert space and it is easy to write formal expressions which
have convergence problems and probably do not make sense.
To write the equations satisfied by physical master fields, we will want to use
concepts like $M^k$ or $\partial M/\p x$ and it would be very useful to know
when these do make sense.
For typical (regulated) actions the operators $\hat M_\alpha$ are bounded and
thus the framework in which to make precise definitions is the theory of
operator algebras.  The algebra $\CF_n$ is a subalgebra of a $C^\ast$ algebra,
the group algebra of the free group on $n$ generators, $F_n$.

This theory provides a construction of the master field given the loop
functional $W_S$ for {\it any}
field theory: the GNS representation.\singer\ %
This construction puts all the information from $W$ in the inner product and it
is not so explicit, but it does demonstrate the existence of the master field
as well as yielding the important result that {\it not} every loop functional
$W$ corresponds to a master field: it must satisfy positivity constraints.\doug

Another representation has been provided by the work of Voiculescu. \voi\ %
The simplest case and a very relevant one for stochastic quantization is the
Gaussian master field.
Let us take $\vev{M_\alpha M_\beta}=\delta_{\alpha \beta}$.
Since we can make linear transformations $M_a=R_a^\alpha M_\alpha$, this is
general.
The variables $M_\alpha$ are free random variables in the sense of Voiculescu:
the expectation of any word is determined by the expectations of the individual
matrices in a way analogous to the construction of a joint distribution of
independent commuting random variables.

These ideas lead to an explicit operator construction of the master field:
\eqn\Gmaster{M_\alpha = a_\alpha + a^*_\alpha}
where $a_\alpha$ and $a^*_\alpha$ are operators defined by the following matrix
elements.
We define the `free Fock space on $S$'
to be a Hilbert space with orthonormal basis vectors labelled by an ordered
list of zero or more elements $\alpha_j\in S$.
\eqn\freefock{\eqalign{
a^*_\alpha \ket{\alpha_n,\alpha_{n-1},\ldots,\alpha_1} &=
\ket{\alpha,\alpha_n,\alpha_{n-1},\ldots,\alpha_1}\hfill\cr
a_\alpha \ket{\alpha_n,\alpha_{n-1},\ldots,\alpha_1} &=
\delta_{\alpha,\alpha_n} \ket{\alpha_{n-1},\ldots,\alpha_1}\hfill\cr
a_\alpha \ket{} = 0.}}
They are similar to bosonic creation and annihilation operators but with two
differences.  First, they are {\it free} -- the product of two operators
associated with $\alpha$ and $\beta\ne \alpha$ satisfies no relations.  Second,
the usual symmetry factor for bosonic harmonic oscillators is absent.
Thus, we have not $[a,a^*]=1$ but instead
\eqn\freecomm{
\sum_\alpha [a_\alpha,a^*_\alpha] = |S|-1+\ket{}~\bra{}.}

The trace is not the standard one (which does not make sense here)
but is defined by
\eqn\freetrace{
\hat\tr A = \bra{}A\ket{}.}
It is a trace in that it satisfies the axioms for a trace, for example
$\hat\tr [A,B]=0$, as we will see.

The simplest proof of \Masdef\ is diagrammatic. \doug\ %
If we expand the product of the $M_{\alpha_i}=a_{\alpha_i} + a_{\alpha_i}^*$ in
$2^k$ terms,
we can associate a planar diagram with each non-zero term.
Consider the example in the figure of
$\vev{\hat\tr \hat M_b \hat M_a^2 \hat M_b\hat M_a^2 }$.
We start with $M_{\alpha_k=a}\ket{}$ which makes the state $\ket{a}$, and we
interpret this as adding a line with the label $a$ to the diagram.
With $M_{\alpha_{k-1}}$, there are in general two possibilities:
we can always act with $a_{\alpha_{k-1}}^*$ to create a new line labeled
$\alpha_{k-1}$, and
if $\alpha_{k-1}=\alpha_k$ we can annihilate the existing line with
$a_{\alpha_{k-1}}$ as well.
The `free' nature of the Fock space precisely reproduces the planarity
constraint on the diagrams.  After applying $M_{\alpha_1}$, we must be left
with
no lines to have a matrix element with $\bra{}$.
This proves \Masdef, and since $\Tr$ is a trace, so is $\hat\tr$.

\fig{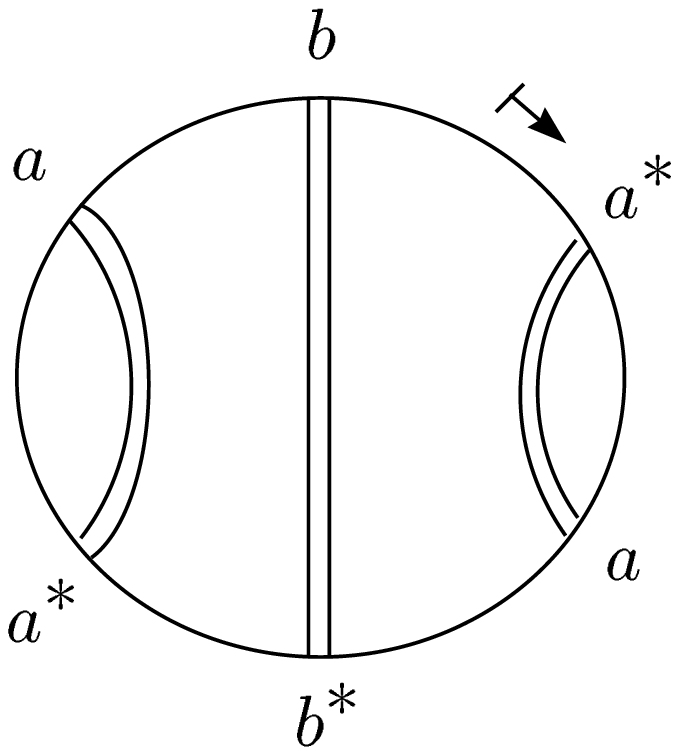}{}


This is a special case of the `free product of distributions,' which given
master fields for the functional integral with action $S_1 + S_2 + \ldots$
allows us to construct master fields for the combined functional integral.
The multi-matrix \Gmaster\ is simply $n$ copies of a single Gaussian master
field acting on separate parts of the free Fock space.
To generalize this we need more general one-matrix master fields.

An explicit master field for a general one-matrix integral
\eqn\omint{\int DM e^{-NV(M)}}
with a similar form is given in \voi:
\eqn\Cmaster{\hat M = a + \sum_n c_n (a^*)^n}
Since the coefficient $c_m$ only affects the
expectations $\hat \Tr \hat M^n$ for $n\ge m$, clearly we can find $c_m$'s
which reproduce any $W$.
The diagrammatic argument works the same way: we associate a diagram to each
term in the expansion of \Masdef, with each power of $a^*$ creating an
additional internal leg.  The free nature of the operators prevents lines of
this diagram from crossing and thus the diagrams are planar.

The coefficients $c_n$ are associated with connected diagrams with $n+1$ legs
in this argument and thus we can write
\eqn\CNfmaster{\hat M = a + f(a^*)}
where $f$ is the generating functional of connected planar diagrams
\eqn\Cfexpand{f(w) = \sum_{n\ge 1} G^{(c)}_{n+1} w^n. }
as introduced in Brezin et. al. \bipz\ %
Although many generalizations of \Cmaster\
to more general matrix models can be made,
this observation (due to R. Gopakumar and D. Gross \gross) will motivate the
particular one we introduce below.
Here, it means that $f(w)$ is given by the results of \bipz: they
define the generating functions of disconnected diagrams
\eqn\defphi{\phi(u) = \sum_{n\ge 0} u^n W(m^n) = W({1\over 1-um})}
and connected diagrams
\eqn\defpsi{\eqalign{\psi(j) &= 1 + \sum_{n\ge 1} G^{(c)}_n j^n\cr
&= 1 + j f(j).}}
The relation is then
\eqn\conrelone{\phi(u) = \psi(j)}
with $j$ defined implicitly by
\eqn\conrelcon{j = u\psi(j).}
In words, every time we attach to an external leg with $u$, we allow for the
possibility of an arbitrary `rainbow diagram' to its left.  The sum of these is
generated by expanding the implicit relation.

For many purposes it is more convenient to work with the resolvent
\eqn\defR{R(z) = W({1\over z-m}) = \sum_{n\ge 0} W(m^n) z^{-1-n}
= z^{-1}\phi(z^{-1})}
and the symbol of the operator $\hat M$
\eqn\defK{
K(j) = j^{-1} + f(j) = j^{-1}\psi(j)}
The relations \defphi-\conrelcon\ are then equivalent to \voi\ %
\eqn\invrel{K(R(z)) = z.}

The free product of several such operators,
giving the master field for a decoupled set of matrix integrals,
can now be constructed simply by taking \voi
\eqn\CNmaster{\hat M_\a = a_\a + \sum_n (c_\a)_n (a_\a^*)^n}
with the operators as above.

A virtue of this representation for present purposes is that the operation
`$+$' between two free master fields (the `additive free convolution' of
Voiculescu) is very simple.
Define a master field by
\eqn\adddef{\vev{M_{1+2}^n} = \vev{\Tr(M_1+M_2)^n}.}

If $M_1$ and $M_2$ are relatively free -- i.e.
they act on independent factors in a free product of
Hilbert spaces -- this is completely determined by $\hat M_1$ and $\hat M_2$.
We can replace
\eqn\addmat{\hat M_1 + \hat M_2 = a_1 + a_2 + f_1(a_1^*,a^*) + f_2(a_2^*,b^*)}
by
\eqn\addmattwo{\hat M_{1+2} = a_1 + f_1(a_1^*,a^*) + f_2(a_1^*,b^*)}
because in each term in the expansion of \Masdef, only a single one of $a_1$
or $a_2$ could contribute.
In diagrammatic language, an appearance of $M_1 + M_2$ produces a sum of terms
where we attach a diagram from $M_1$ or $M_2$, and we do not have to
keep track of which of the two the diagram came from.
The simplest statement of this for a single matrix problem \voi\ is
in terms of the function $f(j)$ of \CNfmaster:
\eqn\addf{f_{1+2}(j)=f_1(j)+f_2(j).}

In stochastic quantization, the noise $\eta(\tau)$ at each time $\tau$ is free
with respect to everything at previous times, and thus this result will apply
to the Langevin equation for the master fields.

A natural generalization to multi-matrix models is to consider representations
of the form
\eqn\Crep{\hat M_\a = a_\a + f_\a(a_1^*,a_2^*,\ldots,a_n^*)}
which we will call a  master field in the `$C$-representation.'
The functions $f_\a$ are general functions depending only on the creation
operators.  Since these operators do not commute, these are functions on a free
algebra and a basis for these is rather large: for example,
\eqn\fbasis{f(a^*,b^*) = f_{\hat a}(a^*) + f_{\hat b}(b^*)
+ f_{{\hat a}b}(a^*) f_{a{\hat b}}(b^*)
+ f_{{\hat b}a}(b^*) f_{b{\hat a}}(a^*) + \ldots .}
As for the one-matrix $c_n$'s, there are enough coefficients in a series
expansion of these functions to reproduce functionals $W_S$ for very general
actions.
(The question of which $W_S$ can be reproduced is non-perturbative.)

{}From the matrix integral, the result \Masdef\ would be computed as a sum of
disconnected diagrams and thus, following the observation of Gopakumar and
Gross,
we reproduce it if $f_\a$ is the generating function for connected
diagrams with a marked external $M_\a$ leg,
with a term $a_{i_1}^*\ldots a_{i_k}^*$ creating internal legs $M_{i_1}\ldots
M_{i_k}$.
In writing the expression \Masdef\ we break the cyclic symmetry of the trace in
the original Green's function.  This allows us to assign each connected
component of a disconnected diagram a unique `first leg,' the first to appear
in reading \Masdef\ from right to left, and thus there are no additional
symmetry factors: each disconnected diagram is
generated once in the expansion.

The result of Brezin et. al. generalizes to any number of matrices.
We again introduce generating functions $\phi(u_\a)$ and $\psi(j_\a)$ of
disconnected and connected diagrams respectively.
The variables $u_\a$ and $j_\a$ are free (non-commuting) and we have expansions
like
\eqn\noncomexp{\eqalign{\phi(u) &= 1 + \sum u_1^n W(m_1^n) + \sum u_2^n
W(m_2^n)
+\ldots + \sum u_1^m u_2^n W(m_1^m m_2^n) + \ldots\cr
&\equiv ``W({1\over 1 - \vec u \vec m})".}}
The analog of \conrelone-\conrelcon\ is
\eqn\conreltwo{\eqalign{&\phi(u) = \psi(j)\cr
&j_\a = u_\a\psi(j).}}
These sum all rainbow diagrams.

The Gaussian two-matrix case is a good illustration:
$\psi(j,k) = 1 + j^2 + k^2$
so $j = y + y(j^2 + k^2)$, $k = z + z(j^2 + k^2)$
and $\phi(y,z) = 1 + y^2 + z^2 + y(j^2+k^2)y + y^2(j^2+k^2) +
y(j^2+k^2)y(j^2+k^2) + \ldots$.  The grading of the terms by word length makes
this expansion well defined.

Can this relation be inverted to compute $\psi(j)$ from $\phi(u)$?
Although there is a theory of functions of free variables (e.g. \nsf), we know
of no general treatment of such problems.
We want to interpret \conreltwo\ the other way around,
solving $j_\a=u_\a\phi(u)$ for $u$ as a function of $j$.
Writing $z_\a=u_\a^{-1}$ and $z_\a=\phi(z^{-1}) j_\a^{-1}$ makes the
inverse relation very similar to the original, as with \invrel.

\expandedversion
A natural analog of \defK\ is to let $K_\a(j)=\psi(j) j_\a^{-1}$.
The symbols for the operators \Crep\ are not quite these but instead are
\eqn\symKn{j_\a^{-1} + [K_\a(j)]_+.}
The operation $[f(j)]_+$ here means to drop any term with a negative power of
any $j_\a$ -- so for $K_\a$ we keep only terms in $\psi(j)$ which end with
$j_\a$.  We can also write a formal functional inverse
$R_\a(u)=u_\a \phi(u)$: if we take $u_\a=K_\a(j)^{-1}$ this will give
$j_a=R_a(u)$.
\fi

Whether we can solve for $\psi$ given $\phi$ in practice,
whether there is any analog of the analyticity properties of $R(z)$
and $K(z)$ in the complex plane, and so forth, is unknown.
Making less formal definitions of these generating functions would also be
valuable to justify the claim (which we believe -- of course this is well
understood in the one matrix case) that the objects under discussion are
well-defined beyond perturbation theory and that all references to diagrams in
the preceding are pictorial only.

The work of \voi\ does not assume that the variables are hermitian matrices and
one can treat gauge theories in this formalism either directly in terms of the
$A_\mu$ as above, or by taking holonomy variables $U=\exp iA$ as free variables
with spectral density on the unit circle in the complex plane.  The latter
would be appropriate for lattice gauge theory or the principal chiral model.

It is worth pointing out a simple application of freeness (in Voiculescu's
sense) in this context.  In the extreme strong limit of lattice gauge theory,
the link integrals are completely independent, and in the large $N$ limit they
go over to free random variables.  The master field is just $U_l\rightarrow
\hat U_l$ with each $\hat U_l$ a generator of a free group and
$\hat\tr O=\delta(O,1)$.  One expects that this is the way confinement will be
realized in any large $N$ gauge theory:\footnote{$^1$}{This argument was
developed in a conversation with D. Minic.}
 gauge-invariant correlators of non-trivial operators such as $\tr
F(x)U_{xy}F(y)U_{yx}$ will fall off exponentially with distance and $F(x)$ and
$F(y)$ will become relatively free.  By working in axial gauge $(x-y)\cdot A=0$
and integrating, it is fairly easy to argue that $\vev{A(x)A(y)}\sim|x-y|$ and
that this will give confinement. \dosch

\newsec{Stochastic Quantization}

This is well treated in the literature \stoquant\ and we just mention a few of
the main points.  The Langevin equation is

\eqn\sto{{\p\over \p\tau} \phi_\alpha(\tau,x) =
-{\delta\over\delta\phi_\alpha(\tau,x)} S[\phi] + \eta_\alpha(\tau,x).}
A quantum expectation is reproduced in terms of the large $\tau$ limit of a
solution $\phi_\alpha(\tau)\rightarrow\tilde\phi_\alpha$ as
\eqn\storep{\vev{\phi_{\alpha_1}\phi_{\alpha_2}\ldots} =
\vev{\tilde\phi_{\alpha_1}\tilde\phi_{\alpha_2}\ldots}_\eta}
where the new average is over Gaussian noise with
\eqn\defnoise{\vev{\eta_\alpha(\tau) \eta_\beta(\tau')} =2\delta_{\alpha
\beta}\delta(\tau-\tau').}
The probability distribution $P[\phi,\tau]$ satisfies the Fokker-Planck
equation
\eqn\fokpl{{\p\over \p\tau}P = \sum_\alpha
{\delta\over\delta\phi_\alpha}\left({\delta S\over\delta\phi_\alpha} P\right)
+ {\delta^2\over\delta\phi_\alpha^2} P}
with large $\tau$ limit $P[\phi]=\exp -S[\phi]$.

Perturbatively one can describe the result as a sum of `stochastic diagrams.'
The stochastic diagrams for $\vev{\phi_1\ldots\phi_n}$ are formed by taking all
tree diagrams contributing to each $\phi_\a$ (one external leg) with sources
$\eta$ (zero or more external legs), and tying these together in all possible
ways with a quadratic $\eta^2$ vertex.
The Gaussian case illustrates the role of the $2$ in \defnoise: the single
diagram has two propagators $e^{-\tau(k^2+m^2)}$ and a vertex.
We integrate the result $\int d\tau$, and the vertex factor $2$ compensates the
extra $1/2$ produced by the integral.

Recently a nice interpretion of the proper time in these diagrams has been made
by Jevicki and Rodrigues and by Ishibashi et. al.
\ref\kawai{A.~Jevicki and J.~Rodrigues, preprint HET-827 and LPTENS-93-52,
hep-th/9312118;
N. Ishibashi, H. Kawai, T. Mogami, R. Nakayama, and N. Sasakura,
preprint KEK-TH-411, hep-th/9409101.}:
it is the analog of the time coordinate $\tau$ on a string world-sheet in
`temporal gauge.'
In this gauge, $\tau$ is defined as the minimum geodesic distance to a boundary
of the diagram.

All loop diagrams come from sewing tree diagrams with $\eta^2$ vertices and
thus we can regulate the theory by regulating the $\vev{\eta \eta}$ correlator.
 This is `stochastic regularization' and is particularly attractive for gauge
theories. \refs{\breit,\bhst}\  The present discussion will however assume that
the theory of interest is already regulated, for example by the lattice.

\newsec{Stochastic Master Fields}

Greensite and Halpern pointed out that when the variables $\eta_\a(\tau)$ are
large $N$ matrices, a `master noise' can be constructed to reproduce \defnoise.
In the present language this is just \Gmaster.
This leaves us with the problem of defining the rest of \sto.

We can define the first order proper time derivative in \sto\ as
the $\epsilon\rightarrow 0$ limit of the difference equation
\eqn\differ{M_\a(\tau+\epsilon)=M_\a(\tau) + \epsilon \dot M_\a(\tau).}
We will assemble $\dot M_\a(\tau)$ as the sum of two terms.
One term is $\eta(\tau)$, which is always free with respect to the rest of the
equation.  We can thus rewrite the equation
\eqn\stosto{\dot M = \eta(\tau)}
using \addmattwo\ as
\eqn\stomult{\dot f_\a(a^*) = 2 a_\a^*.}
In the one-matrix case this is equivalent to
\eqn\stostoK{\dot K(j) = 2 j.}
Using $\p_\tau(K(R(z)))=0$, we can change variables in this to $R(z)$,
and find \refs{\doug,\voi}\ %
\eqn\hopf{\dot R(z) + \p_z R(z)^2=0,}
the Hopf equation of matrix quantum mechanics.
Its general solution is thus most simply written in terms of $K(j)$.

The forcing term in the equation is harder to write.
Let us first do it for the one-matrix case.
It is most simply written in terms of $R(z)$
because we can derive it by varying $M$:
\eqn\vary{\eqalign{\Tr{1\over z-M-\delta M(M)} &=
\Tr{1\over (z-M)^2}\delta M(M) \cr
&= -\p_z \Tr{1\over (z-M)}\delta M(M) \cr
&= -\p_z[\Tr{1\over (z-M)}\delta M(z)]_-.}}
where $[F]_-$ is the projection on the part $F(z) \sim 1/z$ at infinity,
i.e.
\eqn\defFF{[F]_- = {1\over 2\pi i}\oint {dz'\over z-z'} F(z')}
with the contour around the cut in $F(z')$.
Thus we can rewrite
\eqn\stoforce{\dot M = -V'(M)}
as
\eqn\stoforceR{\dot R(z) = \partial_z\left[V'(z)R(z)\right]_-}
The imaginary part $\rho(z)=(1/\pi)\im R(z-i\epsilon)$
is the spectral density of $M$ and
thus \stoforceR\ is essentially the analytic continuation of the first term in
\fokpl.

The full stochastic evolution is just the sum of the two terms.
We can change variables to $R(z)$ and find
\eqn\stoR{\dot R(z) = \partial_z\left[V'(z)R(z)\right]_- - \p_z R(z)^2.}
As $\tau\rightarrow\infty$, the solution approaches a
stationary master field with $\dot R(z)=0$ and (integrating once)
\eqn\usualSD{\left[V'(z)R(z)\right]_- = R(z)^2 + c.}
A solution with $R(z)\sim 1/z$ exists only if $c=0$ and this is the usual
factorized Schwinger-Dyson equation for the one-matrix model.

The new possibility is to change
variables in \stoforceR\ to $K(j)$ to find
\eqn\stoK{\dot K(j) = -\partial_j\left[ j~V'(K(j))\right]_- + 2j}
and at the stationary point,
\eqn\stoKstat{\partial_j\left[ j~V'(K(j))\right]_- = 2j}
which we can integrate to get
\eqn\stoKstatt{j^{-1}\left[ j~V'(K(j))\right]_- = j.}
We can regard it as a precise version of
\eqn\stostat{V'(\hat M) = \eta,}
a classical equation of motion for the master field.

Care must be taken in interpreting \stostat\ %
as a naive treatment of it, for example using it to conclude that we can derive
the spectral density $\rho(\lambda)$ at the saddle point by the simple change
of variables $V'(\lambda)=\chi$ from a semicircular $\rho(\chi)$, is incorrect.
We see that the calculus of \voi\ provides a correct interpretation.

It is interesting to ask whether we can
interpret the equation \stostat\ more directly as
an equation for the operator $\hat M$.
Strictly speaking, at the fixed point $\tau\rightarrow\infty$, we have
not $\dot M=0$ but rather that $\dot M$ is an infinitesimal similarity
transformation (`gauge transformation') of $M$.  We should write
\eqn\stostat{0=\dot M=-V'(M) + \eta + [A,M]}
for an operator $A$ to be determined.

If we compute $V'(M) = \sum k V_k M^{k-1}$ in the $C$-representation,
by defining $M^{k-1}$ as a power of \Cmaster, the result will not be of the
form $a+f(a^*)$.
We need to choose $A$ to fix this.

For $V'(M) = m^2 M$, the time evolution $\dot M = -V'(M)$ is
\eqn\extime{M+\eps\dot M = (1-\eps m^2) a +(1-\eps m^2) \sum_k c_k (a^*)^k.}
This is not in the $C$-representation but we can do a simple
`canonical transformation' $a\rightarrow (1+\eps m^2) a$ and
$a^*\rightarrow (1-\eps m^2)~ a$ to find
\eqn\extimetwo{\dot M - [A,M] = - m^2\sum_k (k+1)~ c_k ~ (a^*)^k.}
This is precisely the time evolution of \stoK.
We see that both $-V'(M)$ and $\eta$ must be regarded as infinitesimal
variations, and in this sense \stostat\ is valid.

More generally, we need a similarity transformation which eliminates higher
powers of $a$.
This is a bit more complicated, as is suggested by the operation
$[j~V'(K(j))]_-$  in \stoKstatt\ which is defined to eliminate positive powers
of $u$, not $j$.
Let $A\sim B$ indicate that two operators differ only in matrix elements
$A_{ij}-B_{ij}$ with $i$ and $j$ less than some finite integer.
Then, $[M,a]\sim [M,a^*]\sim 0$.
We need $[[A,M],a]\sim 0$ and this implies $[A,a]\sim p(M)$ for some function
$p$.  This will produce $[A,M]\sim p'(M)$ and the result (which can be verified
by comparison with \usualSD) is that \stostat\ becomes
\eqn\stoKstatt{V'(K) - {1\over j}p'(K)=j}
where $p'(K)$ is the unique polynomial in $K$ which cancels negative powers of
$j$ from $V'(K)$.

For a multi-matrix model, we have
\eqn\stostopsi{\dot \psi(j) = 2\sum_\alpha j_\alpha^2.}
The forcing term is again simple in terms of
disconnected diagrams:
\eqn\stoforcegen{\p_\tau \vev{\Tr M_{\alpha_1} M_{\alpha_2}\ldots} =
-\sum_n \vev{\Tr M_{\alpha_1} \ldots \p_{\alpha_n}S[M] \ldots } .}
The derivative $\p_\a M_{\alpha_1}\ldots$ here is a cyclic sum of terms
$\delta_{\a,\alpha_i} M_{\alpha_{i+1}}\ldots M_{\alpha_{i-1}}$.
The diagrammatic interpretation is a sum of terms where we attach $k-1$ legs of
$S[M]$ to an existing diagram, leaving one leg.

If we change variables in \stostopsi\ and go to the fixed point,
this will produce the standard
`integrated' Schwinger-Dyson equation:
\eqn\intSD{\sum_\a \vev{{\p S\over \p M_\a}{\p \over \p M_\a}
\Tr M_{\alpha_1} \ldots M_{\alpha_n}}=\sum_\a \vev{~
{\p^2\over \p M_\a^2}  \Tr M_{\alpha_1} \ldots M_{\alpha_n}. }}
In a sense this is the multi-matrix generalization of the change of
variables from
\stostoK\ to \hopf\ which solves the Hopf equation.
In computing this, the relation $j_\a=u_\a\psi$ gives us
$\dot\psi = 2\sum (u_\alpha\phi)^2$ which is the piece of the right hand side
of \intSD\ in which a derivative acts on the first matrix; the rest comes
from writing $\phi(u)=\psi(j)$ but taking into account the variation of $\psi$
in the relation $j_\a=u_\a\psi$.  This must produce the other
insertions of $u_\a\ldots u_\a$ necessary to produce a cyclically symmetric
answer, and we verified this diagrammatically.

\newsec{Comparison with Previous Work}

After the completion of the original version, we discovered the almost
forgotten papers \haan\ and \cls.

In a beautiful and prescient work of Haan \haan, the first precise master
fields for higher dimensional large $N$ matrix models were constructed.  An
explicit construction for the $D$-dimensional Gaussian field theory essentially
identical to \Gmaster\ and \freefock\ was given.
It was shown for the field theory with $\Tr M^4$ interaction (by an argument
which generalizes trivially to any action) that the master field satisfies the
equation
\eqn\sdmast{i{\p S[\hat M]\over \p \hat M_i} + 2\hat\Pi_i\ket{}=0}
with
\eqn\pialg{[\hat\Pi_i,\hat M_j]=-i\delta_{ij}~\ket{}\bra{}.}
(This algebra arose independently in \refs{\jevicki,\halpschw}).
The equation was studied explicitly in $D=0$ and for Gaussian field theory.
(See also \grosspa.)
It was argued that the usual adaptations of the GNS construction for quantum
field theory demonstrate that such a master field exists.

In \cls, Cvitanovic, Lauwers and Scharbach gave the relation \conreltwo\
between the generating functionals of planar disconnected and connected
diagrams $Z[j]$ and $W[J]$
(treated as functions of free variables),
as well as the relation with the generator of 1PI diagrams.
A number of useful results were derived:
one was the following representation of the disconnected planar Green's
functions (their equation (4.5)):
\eqn\clsrep{
\vev{\Tr M_a \ldots M_z} =
\left({\p W[J]\over \p J_a}+{\p \over \p J_a}\right)
\ldots
\left({\p W[J]\over \p J_z}+{\p \over \p J_z}\right) 1,}
where the $J_i$'s are free variables and
\eqn\freeder{{\p \over \p J_i} J_j f(J) = \delta_{ij}f(J).}
If we make the identification
\eqn\fool{a^*_i = J_i;\qquad a_i = {\p \over \p J_i}}
we see that this is identical to \Crep.
They also derived the equation of motion
\eqn\finalblow{{\p \over \p M_i}S\left[
{\p W[J]\over \p J}+{\p \over \p J}\right] 1
= J_i.}
However, they did not realize that these operators could be interpreted as a
master field.

\newsec{Conclusions}

The theory of operator algebras allows one to construct master fields for
general large $N$ field theories.
A particularly useful version of this for special cases can be
taken from the work of Voiculescu.
Gopakumar and Gross have recently explained its relation to conventional
diagrammatic methods: this master field encodes
the generating functions for connected planar diagrams.
Using this observation it is straightforward to adapt Voiculescu's
construction to general large $N$ field theories.
We studied the stochastic equation of Greensite and Halpern for the master
field, and explained the relation of the equation in this representation to the
standard factorized Schwinger-Dyson equations.

Perhaps the most important outcome is that a
number of seemingly disparate approaches to large $N$ have now been connected
in a precise way, and the subject put on a firmer footing.
Along the way we rediscovered the neglected but we believe important
work of \haan\ and \cls.  In particular, \finalblow\ appears to be the simplest
form for the equation of motion determining the master field.

The connection with the work of Voiculescu is recent and many of his results
and derivations
are new for the large $N$ application (such as the derivation of
matrix quantum mechanics of \doug; see \grosspa\ for more examples.)
His concept of freeness is fundamental and can be used in a number of physics
applications -- an example is the Neu-Speicher model of
localization. \ref\nsp{P.~Neu and R.~Speicher, ``Rigorous mean field model for
CPA: Anderson model with free random variables,'' Heidelberg preprint
HD-TVP-94-17, cond-mat/9410064.}\ %
In section 2 we pointed out its relevance for confinement.
More generally,
the relevance of the theory of operator algebras to large $N$ is now clear.

The concept of master field is rather protean and we believe it will be useful
to consider other representations (equivalent up to similarity transformations)
as well, and be able to convert between representations.
We might draw an analogy with the use of
different gauges in analyzing gauge theory to exhibit different properties of
the theory.
The GNS construction directly relates the original Green's functions \vevdef\
to another master field; thus a simple example of a problem in this direction
would be to show that the
relation \conreltwo\ is invertible in practice, determining the master field in
the C-representation given the original Green's functions.
Probably, other valuable representations for multi-matrix and higher
dimensional
models
remain to be discovered.

To test these ideas we need to work with higher dimensional theories.
Even matrix quantum mechanics (with a general potential) qualifies,
since an explicit master field
describes non-equal-time correlators, and thus contains more information than
the existing solution.

At present, we would guess that the question of whether or not a particular
higher dimensional theory is integrable or not has the same answer at finite
$N$ and large $N$.  If so, applications to non-integrable theories such as
higher dimensional gauge theory will require developing approximate methods.

Developing analytic approximations for loop functionals and master fields is an
important open problem, and it is here that we believe the theory of operator
algebras will show its deeper relevance for these problems.
Paradoxically, we did not really use it in the present work: the algebra
defined by \freefock\ was treated in a purely formal and combinatoric way.
In reality the theory of operator algebras is a branch of analysis, and our
hope is that out of it and the physical applications will eventually grow
practical tools which allow representing realistic loop functionals and master
fields with a finite amount of information.

\medskip
We thank R. Gopakumar, D. Gross, M. Halpern, M. Li,
D. Minic, G. Moore, S. Shenker and I. Singer
for useful discussions.  We especially thank M. Halpern for pointing out \haan.

After the completion of the original version
we received the work of R. Gopakumar and D.
Gross \grosspa, where the representation \Crep\ is also proposed.

This research was supported in part by DOE grant DE-FG05-90ER40559, NSF
PHY-9157016 and the A. J. Sloan Foundation.
\listrefs
\end